\documentclass{aa}
\usepackage{aabib99}
\usepackage{graphicx,psfig,amsmath}

\begin{document}

   \thesaurus{02     
              (11.16.1;  
               11.17.3;  
               12.03.2)} 

   \author{E. Hatziminaoglou,
           G. Mathez,
           R. Pell\'o
          }

   \offprints{eva@ast.obs-mip.fr}

   \institute{Observatoire Midi-Pyr\'en\'ees, Laboratoire
d'Astrophysique, UMR 5572, 14 Avenue E. Belin, F-31400 Toulouse, France}

   \date{Received, accepted}
   \authorrunning{Hatziminaoglou et al.}
   \titlerunning{Quasar Candidate Multicolor Selection Technique}

   \title{Quasar Candidate Multicolor Selection Technique: 
a different approach.} 
   \maketitle

   \begin{abstract}

We present a quasar candidate identification technique based on
multicolor photometry. The traditional multi-dimensional method
(2 $\times$ N dimensions, where N is the number of the color-color
diagrams)
is reduced to a one-dimensional technique, which consists in a
standard fitting procedure, where the observed spectral energy
distributions are compared to quasar simulated spectra and stellar
templates. This new multicolor
approach is firstly applied to simulated catalogues and its {\it
efficiency} is examined in various redshift ranges, 
as a function of the filter combination and the available observing 
time for spectroscopy.  We conclude that
this method is better suited than the usual multicolor selection
techniques to quasar identification, especially for high-redshift
quasars. The application of the method to real quasar samples
found in the literature results in an {\it efficiency} comparable to the
one obtained from the use of color-color diagrams. The major advantage
of the new method is the estimation of the {\it photometric} redshift of
quasar candidates, enabling, in almost all cases, 
spectroscopy to be targeted to best suited wavelength ranges.

      \keywords{imaging survey -- photometric redshift -- quasars}

   \end{abstract}


\section{Introduction}
\label{intro}
Nowadays, very high redshift galaxies and quasars are being found. 
However, quasars still offer the opportunity of
a (relatively) easy construction of high-z samples, due to their high
luminosity, quasi-stellar shape (allowing accurate photometry), and to their
emission lines (allowing spectroscopical identification).
Therefore, many large optical quasar surveys are in progress or in
project. Even if high energy emission is quite a common quasar 
property, X-ray samples still need the detection of emission lines in 
the optical domain for the purposes of redshift determination. 
To improve the efficiency of all quasar studies, one 
has to increase both the size and the upper redshift
limit of the sample, and to achieve a maximum
completeness over a redshift range as large as possible.

These surveys are essentially based on the photometric
preselection of quasar
candidates, in order to optimize the efficiency of telescope time.
Preselection makes the completeness questionable both at low
and at high redshift (see discussion in e.g. Miller \& Mitchell, 1998). 
Even multiple color-color diagrams  \cite{hall2}
still reveal a bias in the range $2.2 \le z \le 3$, exactly
where a reversing of evolution is thought to occur, making completeness 
specially important. At low redshift, it has been shown 
that the usual morphologic prerequisite for {\it quasi-stellar} 
candidates results in a noticeable deficiency \cite{wolf}. This trend 
may increase the apparent evolution of quasars.

The present paper describes a joint method both for distinguishing between
quasars and stars/galaxies by their photometry {\it and} for obtaining an 
estimate of the photometric redshift of the quasar candidates as well.
In \S \ref{state} we present the current state of the art in the optical
search for quasars. \S \ref{method} gives a brief description of our
method, which has been first validated with simulations, as shown in \S
\ref{simul}, before being applied to real samples (\S \ref{EIS}). 
The last section of this paper (\S \ref{discuss}) 
contains a summary of our conclusions and a brief discussion.


\section{State of the art in the optical search for quasars}
\label{state}

One of the most common problems in optical surveys is the
selection criteria one has to apply in order to distinguish between
stars and quasars, since both categories of objects have (in most cases)
a point--like morphology. Quasars are spatially resolved only at low
redshifts and only when high quality imaging data are available. 
Selection of QSO candidates among stellar objects from photometric data 
is becoming a standard method.
The UVX technique (UV excess) separates stars from 
quasars with a redshift $z \le 2.2$. This technique is still currently being
used, e.g. in the {\it Chile-UK quasar survey}, 
and the {\it AAT \- 2dF\- QSO Survey}. 

To avoid both the redshift limitation $(z<2.2)$ and the
bias towards blue objects, other equivalent techniques 
are also being applied (for example BRX), with quite a high efficiency 
in other redshift ranges. Many Multicolor Surveys use an increased 
number of filters, especially towards the red, with special care paid to
high-z quasars: examples are found in Boyle et al. (1991), Kennefick 
et al. (1997), Osmer et al. (1998) and the references therein,
Jarvis \& Mac Alpine (1998) and the {\it DPOSS} (Digital POSS II). 
In all these cases, the efficiency never exceeds 35\% ($\simeq$ 3 
candidates selected per one real quasar), depending on magnitude and 
redshift.

For all the selection methods described above, the working space is 
the N $\times$ 2D-manifold of N color-color planes. The locus of
quasars at different redshifts in each of these planes is compared to the 
locus expected for Main Sequence stars. For quasars, the allowed 
multi-color space
is defined through simulated template spectra, set to different redshifts.
Basically all the selection procedures use the positions of candidates
in the multi-color space either to compare with the expected locus for
quasars or to compute the distance to the Main Sequence
(Newberg \& Yanny, 1997; Sloan Digital Sky Survey - Fan, 1999; 
Krisciunas et al., 1998). Quasar candidates are among the objects at the
largest distances with respect to this Main Sequence ``Snake''. 
However the selection of candidates with
quasar spectra is basically a 1D problem: the fitting between
the observed spectral energy distribution (SED) and the equivalent 
one obtained from templates (quasars/stars), all of them computed 
within the same photometric system. Two quasar surveys make use
of such a fit to select candidates and to estimate their photometric 
redshifts. The Calar Alto Deep Imaging Survey ({\it
CADIS}) uses specialized filters and a multicolor classification
algorithm to distinguish between stars, galaxies and quasars, and 
the identification of high-z quasars in particular is based on
the photometric redshift method \cite{wolf}. 
The {\it Large Zenith Telescope}, a 6-meter liquid-mirror telescope, 
will provide photometric redshifts for
more than 1 million of galaxies, thanks to
a series of 40 medium-band filters specially designed, 
and will be able to select quasar candidates quite efficiently. 
For the purposes of this paper, even a reduced set of
broad band photometric data on a given object shall
be considered as a very low resolution spectrum ({\it R}$ \sim$ 1!), which 
could be fitted by synthetic templates according to the standard methods 
commonly used on photometric redshift estimates.
Assumptions on the
morphology of the objects could be avoided in principle, in such a way 
that both point-like and spatially 
resolved sources could be examined through the same pipeline. Furthermore, 
it is interesting to have 
a prior idea on the redshift range of each candidate, in order to perform the
spectroscopic follow up in the suitable wavelength range (visible or IR). 

The software described in the present paper has been developed for the needs
of preparing the quasars/AGN sample to be assembled in the future
VIRMOS survey. It can however be applied to all multicolor catalogues.
The forthcoming VIRMOS spectroscopic galaxy survey will consist in 2 surveys:
a deep one and a shallow one. 
The VIRMOS-SHALLOW spectroscopic sample (statistically 1/4 to 1/3 complete to
$I\leq 22.5$, no other preselection) will ensure random completeness,
without any color or morphology preselection. 
The VIRMOS-DEEP survey (complete for all 
objects with $22.5<I\leq 24$) will provide the opportunity
of assembling a quasar sample avoiding the drawbacks of any preselection,
therefore basically free of the usual biases (colors, redshift etc).
So far, very few such samples are available: three examples come from the CFRS 
(I=22.5, Schade et al., 1995), the Faint Galaxy Redshift Survey 
(B=24, Glazebrook et al., 1995) and more recently Cohen et al. (1999) 
to R=24 in the HDF, involving, however, very few quasars. 
We are examining various possibilities of assembling a third quasar
sample in the limits of the VIRMOS-SHALLOW survey, 
trying to get as much as possible of the spectra of the several thousands of
quasars present in this field by selecting photometric candidates.
Comparing the three VIRMOS samples should provide a unique opportunity of
examining the biases introduced by a photometric and/or morphological
preselection on the quasar samples.


\section{Description of the method}
\label{method}

Our method used here is closely based on {\it hyperz}, 
a public photometric redshift code 
presently under development at the Observatoire Midi-Pyr\'en\'ees. Details on
{\it hyperz} will be given in a forthcoming paper (Bolzonella et al.,
2000; see also Miralles \& Pell\'o, 1998; Pell\'o et al., 1999). 
It is basically a standard SED fitting procedure: the observed 
photometric SED of a given object, obtained through
$n$ filters, is compared to the SED computed for a set of template spectra.
The aim is to find the best fit between the observed and the model photometry
through a standard $\chi^2$ minimization procedure: 

$$\chi^2= \sum_{i=1}^n \frac{(F_{obs}^i-F_{mod}^i(z))^2} {\sigma_i^2} $$

where $F_{obs}^i$ and $F_{mod}^i$ are respectively the observed and the 
template 
fluxes in the $i$ band, and $\sigma_i$ is the error on the observed flux in this
band. Fluxes are normalized to an arbitrary reference filter.
This is a quite general method. The photometric redshift is the redshift $z$ 
corresponding to the minimum $\chi^2$ value. As expected, the accuracy 
on the photometric redshift determination
depends strongly on the set of filters used (number and wavelength 
coverage); this is illustrated in the next paragraph.

In order to apply this method to quasars, the template set is made by series of 
simulated quasar spectra. These spectra have been constructed
by varying the slope of the power-law spectra (spectral index) 
in the optical (between 0.0 and 1.0, in 3 steps) while keeping the
ultraviolet index, $a_{UV}$, constant at 1.76 (a value compatible with 
Wang et al., 1998).
Emission lines are included (Ly$_\alpha$, Ly$_\beta$, CIII, CIV, MgII,
SiIV, H$_\alpha$, H$_\beta$ and H$_\gamma$) with gaussian profiles and
typical intensities (e.g. Peterson, 1997), as well as the small blue
bump, centered at 3000 \AA. The redshift value is set between 0 and 7.
Totally, there are 213 quasar spectra, with 71 redshift steps (d$z=0.1$), 
with 3 simulated spectra on each step, obtained by varying the optical 
spectral index. 
The Ly$_\alpha$ forest has been modelized according to Madau (1995), 
while no reddening has been included.
We also included the stellar library of Pickles (1998), which
contains 108 spectra of stellar types ranging from O to M, including
dwarfs and giant stars. Additionally, we included 22 white dwarf and
carbon stars spectra, which are known to contaminate the quasar
candidates catalogues.

When applying the method to a given sample,
the discrimination between stars and quasars is performed in two steps. 
A first distinction between stars and quasars is made by identifying as
star (quasars) the objects that show a better fit to one of the stellar
(quasar) spectra. 
A further selection on the remaining sample can be made, based on 
the value of the reduced $\chi^2$, with stars identified as objects excluded 
as quasars at the 95\% or 99\%  confidence levels. In the present paper,
the terms ``$\chi^2$ selection'' 
or ``no $\chi^2$ selection'' will apply for such a 
distinction. When an object is identified as quasar candidate, a
photometric redshift estimation is automatically provided. 
The number of degrees of freedom, $\nu$, is equal to the 
number of filters minus 2: we lose one degree by normalizing the flux to a
reference filter and one degree because of the ``best fit'' procedure.
In this way, the separation between stars
and quasars is reduced to a one--dimensional fit between the 
data (the observed SED) and the spectral templates.

As expected, and as confirmed by the simulations, the quality of the 
quasar/star discrimination depends on the filter 
combination, as it happens for the photometric redshift estimate in the
case of galaxies.
Not only the number of filters is important, but also the wavelength range 
covered.
Another important point is the width of the filters which must be suited
to the width of quasar spectral features.  A realistic
determination of the photometric errors is the key to a successful
issue for this method.


\section{Simulations}
\label{simul}

The first test of this method has been done on mixed
(quasars + stars) simulated catalogues. In this paragraph we intend to
show the importance of the set of filters used for optimizing the
results. First, let us define an arbitrary and useful quantity 
which is aimed to quantify the efficiency of this procedure.
Let $N_c$ be the number of quasar candidates resulting from
the identification technique, $N_f$ the number of actual quasars found 
as candidates (real quasars effectively selected), and $N_e$ the number 
of expected quasars. In principle, $N_e$ is a quantity which can be
estimated from previous surveys. We define the ``{\it efficiency}'' of the
method as follows: \\
{\it efficiency}={\it completeness} $\times$ {\it confirmation rate},\\
where {\it completeness}$={N_f}$/${N_e}$ and {\it confirmation rate}
$={N_f}$/${N_c}$. 
Obviously, both the {\it completeness} and the {\it confirmation rate} 
take values in the interval [0,1] and so does the {\it efficiency}.

The number of stars in our simulated catalogues was given by the Galaxy
model of Robin et al., 1995 (R95). The star counts correspond to 
the six Deep Multicolor Survey (hereafter DMS) fields, since we primarily 
tested our results with the results of this survey. 
The fields with their surface (in square arc
minutes) and the estimated star numbers are given in table \ref{gal_lat}. 
Quasar numbers were based on previous surveys (Hartwick \& Schade,
1990, hereafter HS90). Catalogues contain $\sim$100 quasars 
and $\sim$3000 stars, with a
magnitude limit of $m_b$=22.2 and photometric errors scaling with
magnitude, with typical values of the order of 0.1 magnitudes. 
Quasars are uniformly distributed in redshift, in order to
obtain reasonable statistical results at all redshifts.

\begin{table}[ht]
\caption{The six DMS fields, galactic coordinates, surfaces and estimated
star numbers.}
\label{gal_lat}
\begin{tabular}{lc|cc}
$l$&$b$&surface (sq') &stars\\
\hline
250&47&286&219\\
129&-63&516&263\\
77&35&552&826\\
337&57&561&581\\
52&-39&537&922\\
68&-51&537&505\\
\end{tabular}
\end{table}

\begin{figure*}[ht]
\centerline{
\psfig{figure=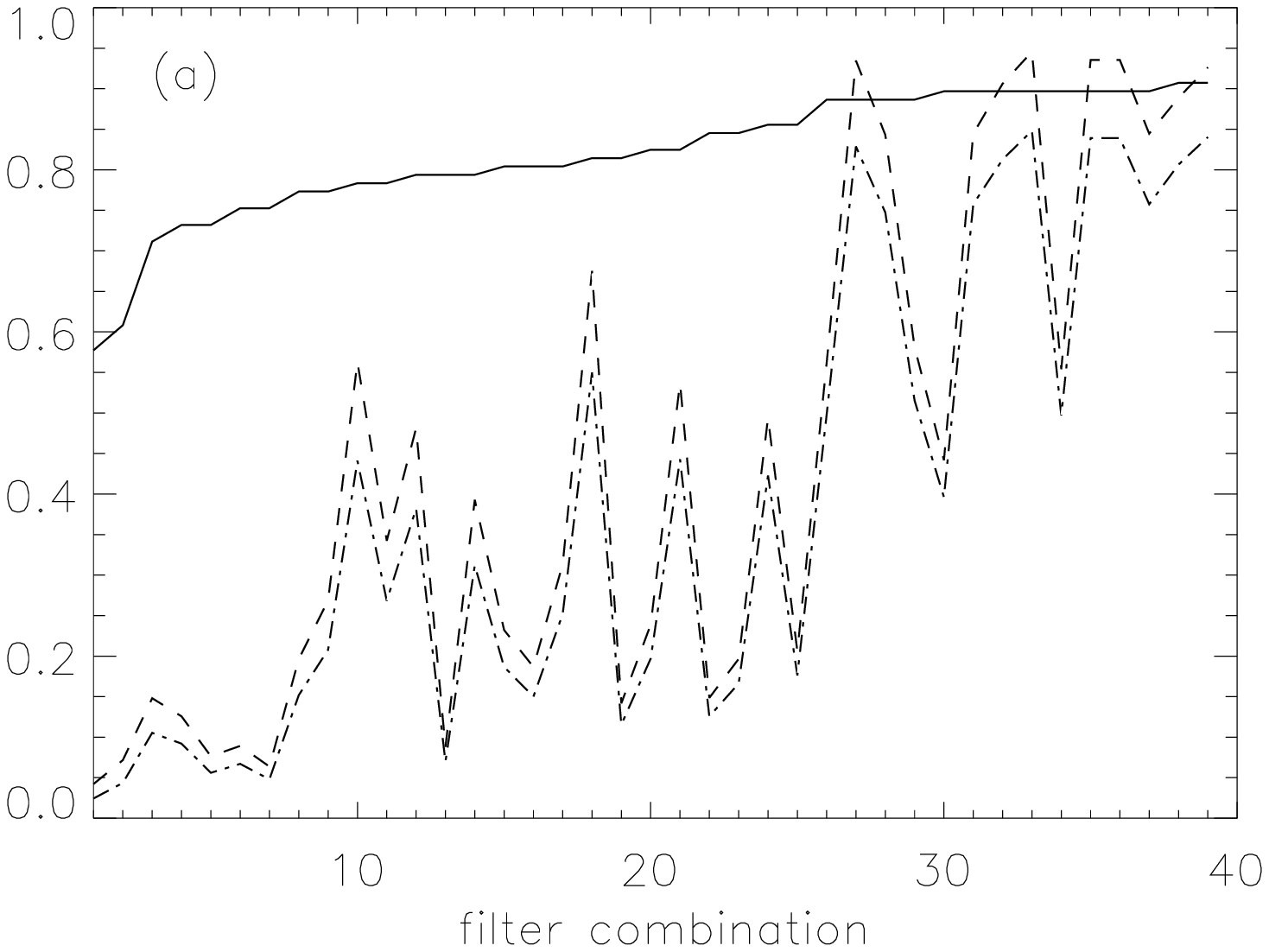,width=7cm,height=6cm}
\psfig{figure=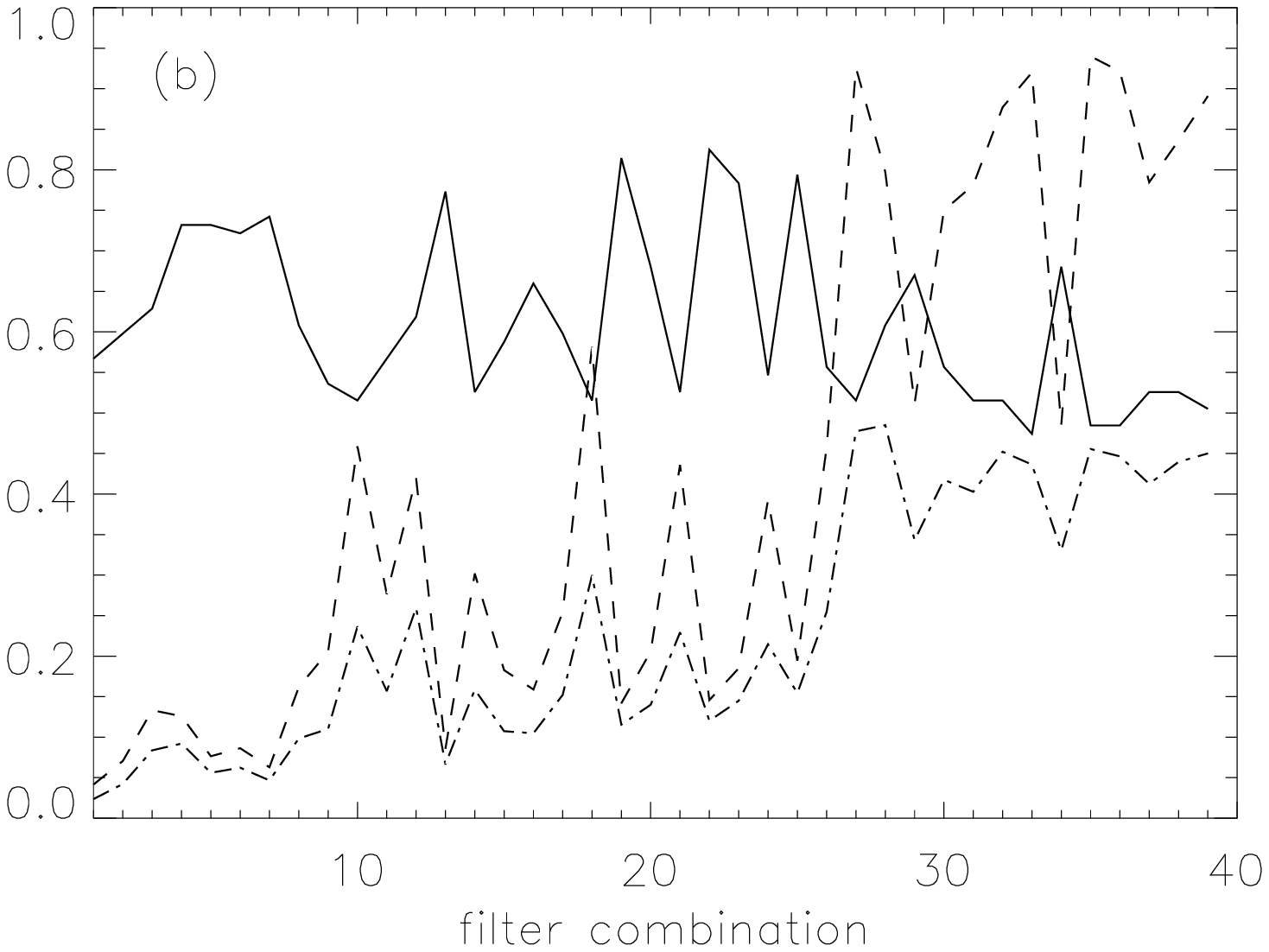,width=7cm,height=6cm}}
\caption{{\it completeness} (solid line), {\it confirmation rate}
(dashed line) and {\it efficiency} (dash-dotted line) for 40 filter
combinations, a) without any $\chi^2$ selection and b) with selection at
a 99\% confidence level.}
\label{chi2sel}
\end{figure*}

Photometry in broad-band filters has been simulated, covering the
wavelength range between $\sim$ 3000 \AA \, and $\sim$ 20000 \AA: U, B, V,
R, I, H, and K. Broad-band filters have been used for practical reasons,
in particular because they directly apply to the VIRMOS survey,
which will make use of
wide-band filters only, although narrow-band filters would give
better results (Peri, Iovino \& Hickson, 1997). 
40 different filter combinations
have been tested in the aim of selecting quasar candidates. Fig.
\ref{chi2sel}a and b illustrate the {\it completeness} (solid line),
{\it confirmation rate} (dashed line) and {\it efficiency} (dash-dotted
line) for these 40 combinations, without any $\chi^2$ selection and with 
selection at 99\% confidence level, respectively. The
filter combinations in fig. \ref{chi2sel}a 
have been classified by increasing {\it completeness}. The same order 
has been used in fig. \ref{chi2sel}b, to demonstrate the
effect of a $\chi^2$ selection on the {\it completeness} and the other
parameters. Table \ref{filter_comb} shows the filter combination 
which corresponds to the abscissa values. 

\begin{table*}
\caption{Filter combinations used in the fig. \ref{chi2sel}a and 
\ref{chi2sel}b, classified by increasing {\it completeness} in the case 
of no $\chi^2$ selection.}
\label{filter_comb}
\begin{flushleft}
\begin{tabular}{lc||cc||cc||cc||cc}
\hline
\hline\noalign{\smallskip}
No&filters&No&filters&No&filters&No&filters&No&filters\\
\noalign{\smallskip}
\hline
\hline
1&VRI&2&VRH&3&BRH&4&UBV&5&BRI\\
6&BVRI&7&BVR&8&VIK&9&BVRIH&10&BRIHK\\
11&VRIHK&12&BRHK&13&BVI&14&BRIK&15&VRIK\\
16&BVK&17&RIHK&18&BVRIHK&19&UBR&20&UBH\\
21&BVRIK&22&UBVR&23&UBRI&24&BVIK&25&UBVRI\\
26&UBVRIK&27&UBVRIH&28&UBRHK&29&UBHK&30&UBVRK\\
31&UVIK&32&UVRIK&33&UBVRIHK&34&UVK&35&UBRIHK\\
36&UBVIHK&37&UBVRHK&38&UVIHK&39&UVRIHK&40&UBRIK\\
\hline
\hline
\end{tabular}
\end{flushleft}
\end{table*}

A closer look at fig. \ref{chi2sel} and Table \ref{filter_comb} reveals that
the filter combination is, most of the time, more important than the
number of filters. In general, filters like U or the infrared filters 
H and/or K are
very useful: they add much photometric information constraining the number
of candidates. Neighboring filters like B, V, R, when used all together,
decrease the {\it efficiency} because, as mentioned above, they increase the
number of the degrees of freedom without adding any essential
information. A comparison of the two figures reveals several
trends. The {\it completeness} is almost not influenced by a
$\chi^2$ selection in the case of 3-filter combinations (such as UBV,
BRI, BVR, BVI, or UBR). It is, however, strongly influenced by a
$\chi^2$ selection when the
filter U is present, in sets of 4 or more filters. Note that this
quantity decreases significantly , when a $\chi^2$ selection is imposed,
for filter combinations with index
higher than 26, which are all sets containing the U-filter.
Furthermore, the {\it confirmation rate} is generally increasing with
increasing number of filters.
The decision whether to make a $\chi^2$ selection or not
depends on the available time for spectroscopy and the scientific 
objectives of the project. 

For this purpose, we made an approximative estimation of the time needed for
making the spectroscopy in order to cover an area of 1 square degree, 
$T_{sp}$, for the same 40 
filter combinations, given a detector area, $S$, a magnitude limit,
$m_b$, a quasar candidate density, $n_q = N_{cand}/S$ (depending on this 
limit as well as on the filter combination),
a number of slits, $G$ and an exposure
time, $dT_{sp}$, with and without any $\chi^2$ selection. We calculate 
$T_{sp}$ as:
$$T_{sp}=\left\lbrace 1+int\left(\frac{N_{cand}}{G}\right)\right
\rbrace \times \frac{dT_{sp}}{S}$$ 

\begin{figure}[ht]
\centerline{
\psfig{figure=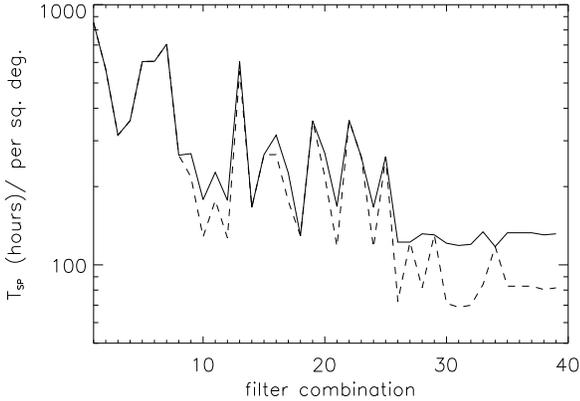,height=6cm}}
\caption{Spectroscopy time (plotted on a logarithmic axis)
per square degree versus filter combination,
without any $\chi^2$ preselection (solid line) and with a 99\%
confidence level selection (dashed line).} 
\label{teltime}
\end{figure}

Fig. \ref{teltime} illustrates $T_{sp}$ (in hours) for $m_b=22.2$,
S=0.06 square degrees, G=10 and $dT_{sp}=$3 hours, without any 
preselection (solid line) and with a $\chi^2$ selection (dashed line).
Note that the y-axis is logarithmic.
As expected, there is an anticorrelation between $T_{sp}$ and the {\it
confirmation rate}, as it appears from the comparison of fig. 
\ref{chi2sel} and \ref{teltime}. The highest the {\it confirmation
rate}, the less time is required.

\subsection{Determination of the Photometric Redshifts}
\label{zphotsim}

\begin{figure*}[ht]
\centerline{
\psfig{figure=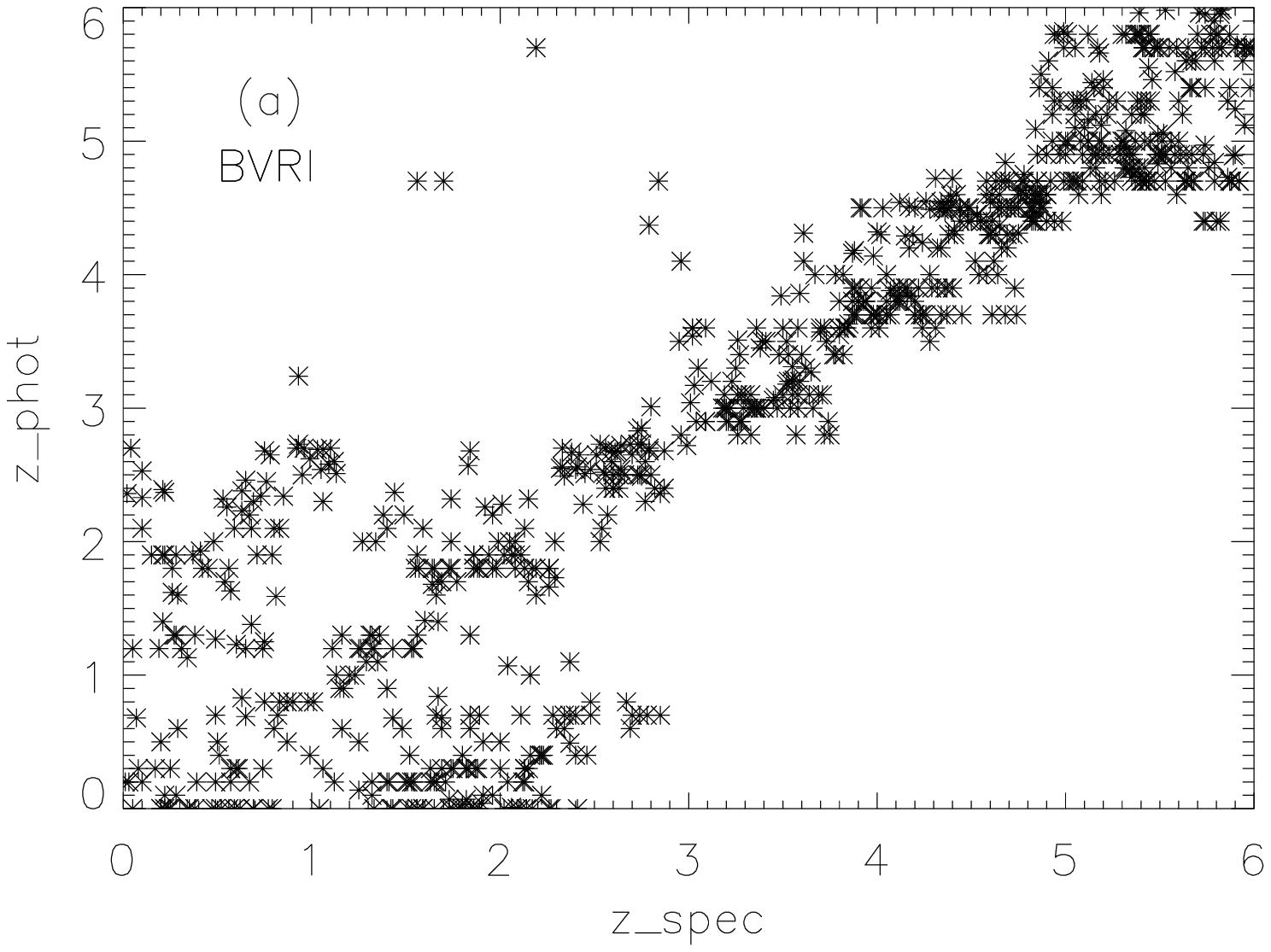,height=7cm,width=8cm}
\psfig{figure=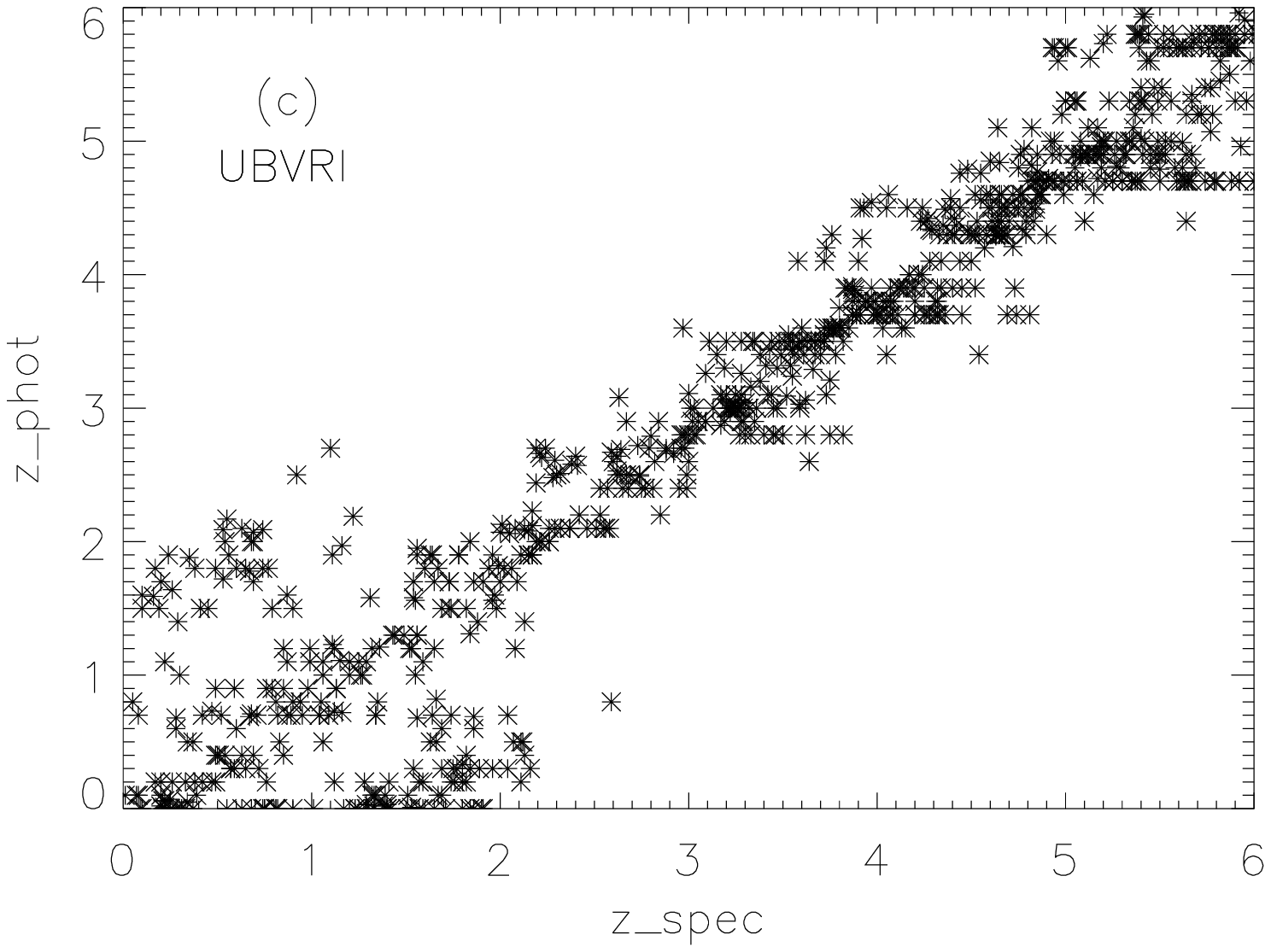,height=7cm,width=8cm}}
\vskip 0.2cm
\centerline{
\psfig{figure=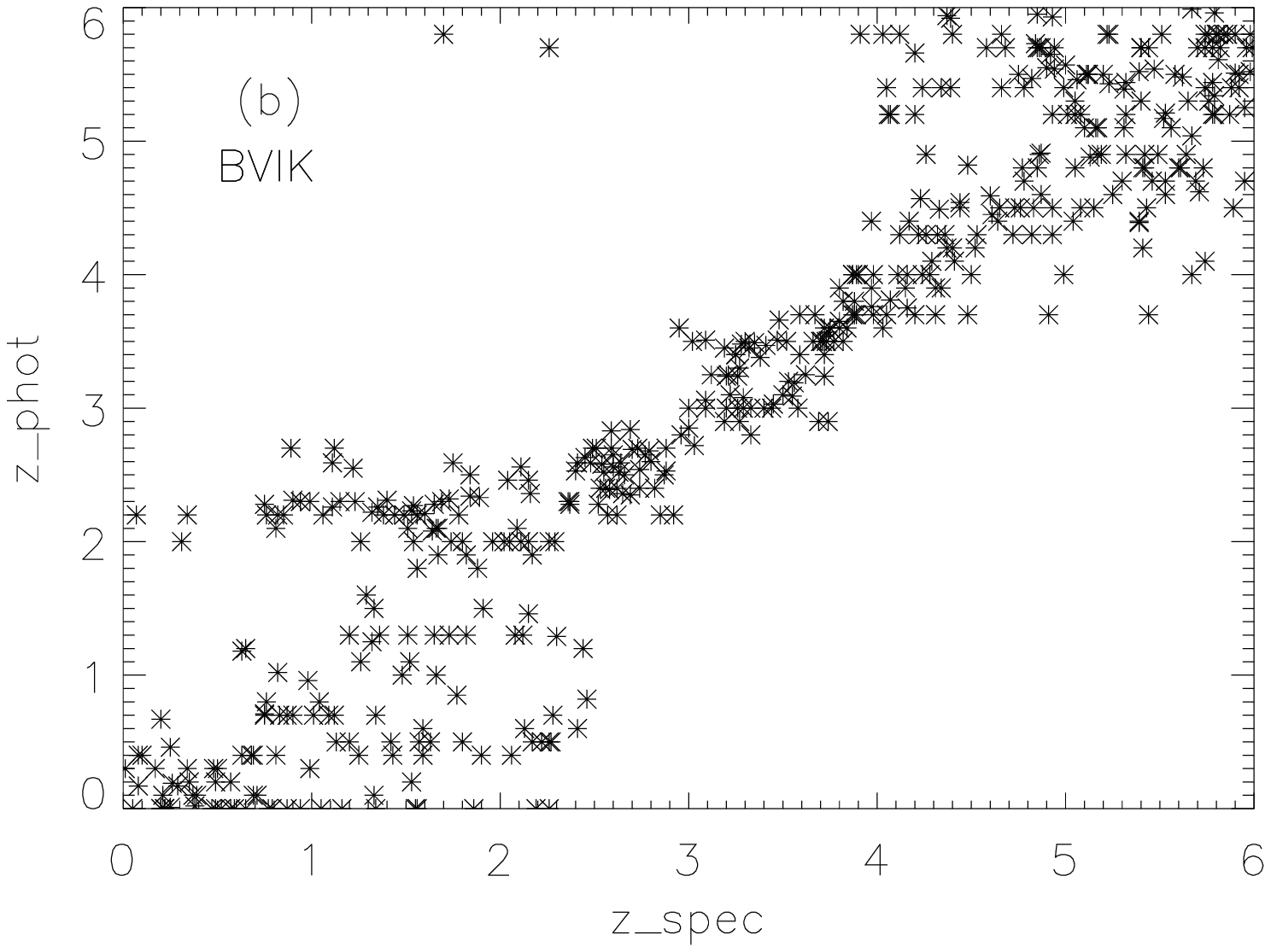,height=7cm,width=8cm}
\psfig{figure=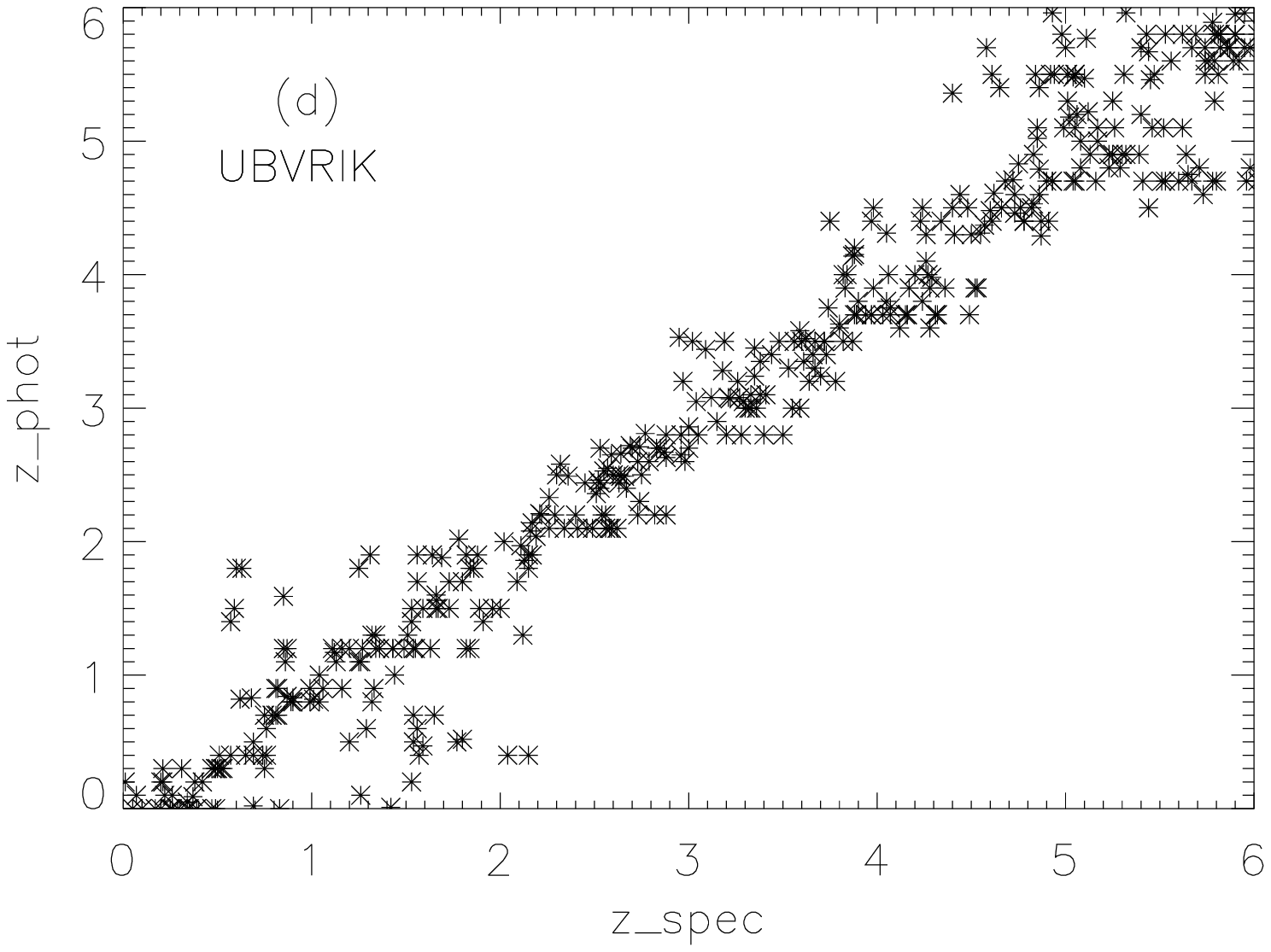,height=7cm,width=8cm}}
\caption{Photometric redshift versus model redshift for a simulated
catalogue of 1540 quasars and 4 different
filter combinations: BVRI (a), BVIK (b), UBVRI (c) and UBVRIK (d).}
\label{zphotsimps}
\end{figure*}

The problem of the quasar photometric redshift (hereafter $z_{phot}$)
estimation is at least as
complex as the one for galaxies. Quasar spectral features (mainly 
emission lines) are rather narrow and the overall shape of the 
ultra-violet and optical continuum is a single power low, while the
filters commonly used for photometry are wide-band filters, not very
suitable for the quasar characteristics. All the above are in the origin
of the dispersed values of the estimated quasar $z_{phot}$,
when compared to the spectroscopic ones, especially for quasars with
redshifts lower than $\sim 2$. The filter combination is very important
for the accurate determination of the $z_{phot}$. Figure
\ref{zphotsimps} shows the photometric redshift versus the model
redshift, for a catalogue of 1540 simulated quasars, uniformly
distributed over the redshift range [0,6], and for 4
different filter combinations. From
top to bottom and left to right the following filter combinations 
are illustrated: BVRI (a), BVIK (b), UBVRI (c), UBVRIK (d).

The use of the filter set BVRI is clearly a bad choice: these four
filters cover a relatively short wavelength range and their effective
wavelengths lie close to each other. The $z_{phot}$ in this case is
poorly determined
for $z \le 3$, while there is also a large dispersion
at $z \ge 5$. If we replace the R-filter by the K-filter (figure
\ref{zphotsimps}b), the dispersion
becomes more important at redshifts higher than 4, since this is the
redshift at which Ly$_{\alpha}$ should enter the R-band, but it
decreases the dispersion at redshifts lower than 3, due to the very
large wavelength range covered.  Adding the filter U to the
BVRI combination decreases the dispersion at both high and low
redshifts (figure \ref{zphotsimps}c), 
but one can still feel the absence of an infrared filter,
whose importance is seen in figure \ref{zphotsimps}d. The periodic
accumulations of points at different redshifts are due to the passage of
Ly$_{\alpha}$ from one filter to the other. As a general rule, the
determination of the photometric redshifts should be rather accurate in
the redshift range [3,5] even without the U or an infrared filter, but
their presence is crucial for all other redshift ranges.

The quality of the $z_{phot}$ determination is likely to improve if real
quasar spectra are used. However, this implies that a spectral library
is required, containing some hundreds of quasar spectra, with redshifts
spanning from 0 to $\sim 5$ and with a spectral coverage spanning form the
ultra-violet to the infrared.

\subsection{C, CR and E as a function of redshift}
\label{Evsz}

The {\it completeness}, the {\it confirmation rate} and, therefore, the
{\it efficiency} of the method depend on the redshift. 
The demonstration of this dependence is easy to show for the {\it
completeness}, but not for the {\it confirmation rate}. The {\it
completeness} within redshift bins has the same definition as for the
entire sample: the ratio between the number of real quasars with
spectroscopic redshifts (hereafter $z_{spec}$) 
within the examined bin, identified
as candidates even if their $z_{phot}$ is significantly different from
their $z_{spec}$, over the estimated total number of quasars. 

\begin{figure}[ht]
\centerline{
\psfig{figure=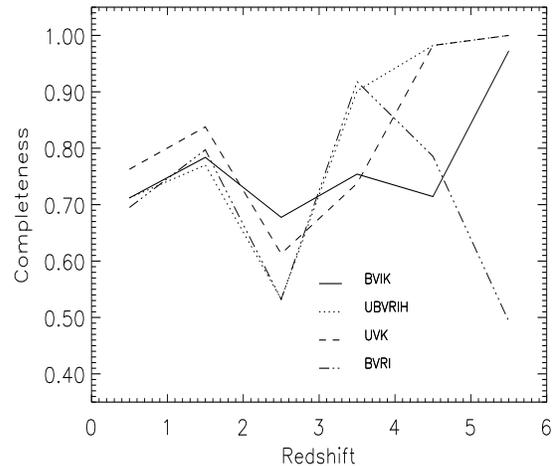,height=7cm,width=8cm}}
\caption{Completeness versus redshifts for 4 different filter sets: ---
BVIK, $\cdot \cdot \cdot$ UBVRIH, - - - UVK, -$\cdot \cdot \cdot$- BVRI.}
\label{cvsz}
\end{figure}

Figure \ref{cvsz} illustrates the {\it completeness} versus the redshift,
without $\chi^2$ selection,
for 4 different filter sets: the solid, dotted, dashed and dashed-dotted
lines correspond to BVIK, UBVRIH, UVK and BVRI, respectively. 
The {\it completeness} over the whole
redshift range for this 4 filter combinations is in all cases more than
85\%, as shown in figure \ref{chi2sel}a.

Figure \ref{cvsz} reveals that, independently of the filter 
combination, the {\it completeness} is always lower in the redshift
range [2,3] due, as already mentioned, to the stellar--like colors of
quasars. This fact could certainly increases the number of quasar
candidates within this range but it could also be a possible 
source of incompleteness,
since some quasars will inevitably be classified as stars. 
Furthermore, different filter combinations offer a
different {\it completeness} at different redshifts. The filter set
BVIK assures a {\it completeness} of over 70\% in the critical redshift
range [2,3], which however remains low up to $z\sim5.5$. UVK offers the
possibility of a {\it completeness} of $\sim$80\% for $z<2$, while
UBVRIH results give a {\it completeness} over 90\% for high-$z$ quasar
samples ($z > 3$).

Things become more complicated if one tries to make an equivalent
evaluation of the {\it confirmation rate} within redshift bins.
This time, the {\it confirmation rate} is the ratio between the number
of real quasars with $z_{phot}$ in the examined redshift bin, over the
number of quasar candidates (quasars + possibly stars) with $z_{phot}$
that belong to this particular redshift interval. Obviously, this
definition depends on the accuracy of the determination of the
$z_{phot}$ as well as on the width of the redshift bins. Therefore,
the dependence of the confirmation rate versus the redshift
can be evaluated qualitatively but not quantitatively, and
so it goes for the {\it efficiency} as well.

If the final aim of a survey is assembling a
quasar sample with redshifts within a restrained interval, all
candidates with a $z_{phot}$ that lies outside this interval will, in
principle, be rejected. In the more realistic case of an all redshift
survey, the most interesting strategy is to simply optimize 
the spectroscopy towards
the optical or the infrared wavelengths. The spectroscopy of quasars 
with redshifts up to $\sim$3 is much more interesting in the optical
band, while higher redshift quasars should be spectroscopically observed
in the infrared, since most of their emission lines are shifted to these
wavelengths. The simulations have shown that, although the determination
of quasar $z_{phot}$ is sometimes dispersed at $z \le 2.5$ and $z \ge
4.5$ (figure \ref{zphotsimps}), this effect hardly translates into
high-$z$ quasars ($z > 4$) being wrongly identified as low-$z$ ones ($z <
3$) or vice versa.

\section{Applying the method to real samples}
\label{EIS}

The method presented above has been tested on a series of 
samples found in the literature, giving rather satisfactory results.
Depending on the sample and the available information, the {\it
confirmation rate} and/or the {\it completeness} have been calculated.

\subsection{The Deep Multicolor Survey}

\begin{figure}[ht]
\centerline{
\psfig{figure=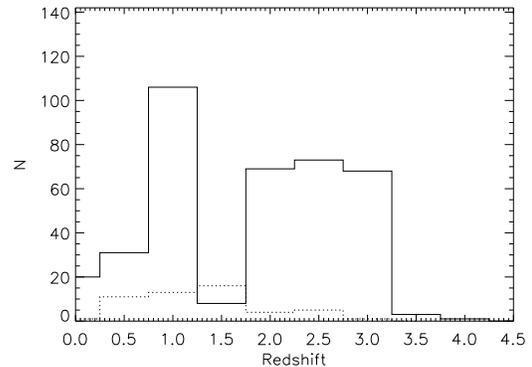,height=5cm}}
\caption{Estimated redshift distribution of the photometrically selected
candidates (solid line), spectroscopically determined sample of
51 DMS quasars (dotted line).} 
\label{histo}
\end{figure}

The Deep Multicolor Survey
is an imaging survey covering 6 fields on a total area of 0.83 square 
degrees on the northern sky at high galactic latitude, carried out at 
the 4m Mayall telescope of the 
Kitt Peak National Observatory. 6 filters have been used, covering the
wavelength range from 3000 to 10000 \AA: U, B, V, R', I75 and I86, and a
photometric catalogue of $\sim$ 21000 stellar--like objects has been 
assembled, with the aim of conducting a multicolor quasar search (Osmer
et al. (1998) and the references therein).
137 stars, 49 compact narrow emission line galaxies (hereafter CNELGs),
and 54 quasars with redshifts spanning from 0.3 to 4.3 
have already been spectroscopically identified.  These objects 
were the 240 quasar candidates selected by Hall et al. 
according to their position on color-color plots, a high fraction of
them (194 objects) selected by the UVX technique (Hall et al., 1996).
This selection favored quasars with
redshifts less than 2.2. The redshift distribution of these objects can
be seen in fig. \ref{histo}, where the spectroscopic sample is 
plotted by a dotted line. 

For our purposes, we cut the catalogue at B=22.3, a limit at which the
sample is almost 100\% complete and only marginally contaminated ($\sim$
8\%) by galaxies, and we only used objects for which photometry 
was available in all 6 filters. Hence, 3 of the spectroscopically
confirmed quasars were excluded, 
and we ended up with a catalogue of 3720 stellar-like objects. 
For the magnitude limit and the area covered by the survey,
approximately 106$^{+27}_{-22}$ quasars are expected to be found among 
the objects of the catalogue (HS90), resulting a
{\it completeness} of at least 80\% for the usual multicolor selection
techniques (Hall, private communication) and $\sim $88\% for our method.

\begin{table*}
\caption{Quasars candidates for DMS, HDF$_{Bright}$ and Parkes selected
by the z$_{phot}$ method.}
\label{real_samples}
\begin{flushleft}
\begin{tabular}{lcccccccccccc}
\hline
\hline\noalign{\smallskip}
 \multicolumn{1}{l}{}
& \multicolumn{1}{c}{}
& \multicolumn{1}{c}{}
& \multicolumn{1}{c}{}
& \multicolumn{2}{c}{Stars} 
& \multicolumn{5}{c}{Quasars}\\
Survey&Surface&Mag.&Filters&&&&&Candidates&&\\
&covered&limit&&exp&cand&exp&95\%&99\%&no $\chi^2$&conf\\
\noalign{\smallskip}
\hline
\hline
\\
DMS&0.83sd&B=22.3&U,B,V,R',&$3316\pm 60$&3332&$106^{+27}_{-22}$&295&348&388&51\\
&&&I75,I86&(R95)&&(HS90)&(40)&(45)&(45)&\\
HDF$_B$&1.00sd&B=21.0&B,V,R&--&&30&26&30&30&30\\
Parkes&&H=19.6&B,V,R,I,&--&&129&99&107&125&129\\
&&&J,H,K&&&&&&&\\
\\
\hline
\hline
\end{tabular}
\end{flushleft}
\end{table*}

Table \ref{real_samples} summarizes the characteristics of the 
different catalogues examined in this section: surface
covered, magnitude limits, filters 
used, number of quasars, and number of candidates for different
confidence levels. The expected
number of stars on the DMS field has been computed from
R95. In the case of the DMS field, the number of spectroscopically 
confirmed quasars is also given between brackets.

Among the 3720 objects in this catalogue, 
we identified 3332 as stars because their photometry shows a 
better match to the stellar templates.
With our method, 6 of the 51 spectroscopically confirmed quasars
(DMS0059-0056, DMS1358-0054, DMS1714+5012, DMS1714+5003, DMS1714-4959
and DMS1358-0055) are misclassified as stars.  Only 7 of the 97 
B $\le 22.3$ candidates, spectroscopically identified as stars,
but all 34 B $\le 22.3$ CNELGs belong to our candidate list,
if no $\chi^2$ selection is made. At a 99\% confidence
level, 25 CNELGs are classified as quasar candidates.

To summarize, we find 388 quasar candidates for 106 actual quasars, with
an expected {\it completeness} of 45/51=0.88 and an expected {\it
confirmation rate} of (106 $\times$ 0.88)/388=0.24, if no $\chi^2$
selection is made. The equivalent values with 99\% confidence level 
selection are 0.88 and 0.27, respectively, for 348
quasar candidates (see also Table \ref{efficiency}).

Fig. \ref{histo} displays with a solid line the estimated redshift 
distribution of our 388 quasar candidates. 
The accumulation of 
identifications around $z \sim 1$ is due to the CNELGs, with
$z_{spec}$ in the range [0.1,0.8] typically, identified by
our procedure as quasar candidates with $z_{phot} \sim1$. The broad
distribution at $z \in [2.0,3.0]$ is probably due to the stellar--like 
colors of quasars at these redshifts, which makes the distinction
between stars and quasars particularly difficult by photometric means,
increasing the number of quasar candidates, but it also corresponds to
the expected maximum of the quasar space distribution.
For comparison, we plot the redshift distribution of the 51
spectroscopically confirmed quasars (dotted line). 

The above results are summarized in the first 6 columns of Table
\ref{efficiency}, where the values of the {\it completeness} (C),
the {\it confirmation rate} (CR) and the {\it efficiency} (E)
are shown in columns 3 and 4. The values predicted by the
simulations are also given for comparison in columns 5 and 6,
using the same filter combinations as for the real data, and
keeping the filter responses as close as possible to those actually
used.
Column 2 gives the same quantities for the DMS and the usual multicolor
selection techniques. Note that the values given for the {\it
completeness} and the {\it efficiency} are the lower estimations. 
The two methods give comparable values for all
3 quantities (C, CR and E), with a slightly better efficiency for the
multicolor technique when a $\chi^2$ criterion is imposed.

\begin{table*}[ht]
\caption{{\it Completeness} (C), {\it confirmation rate} (CR) and {\it
efficiency} (E) for the samples DMS, HDF$_B$ and Parkes and for the
simulated catalogues, with a $\chi^2$
selection at 99\% confidence level and without any $\chi^2$
selection. In column 2 we give the same quantities for the DMS sample
and for the usual multicolor selection techniques.}
\label{efficiency}
\begin{tabular}{l|ccccc|cccccccc}
\hline
\hline\noalign{\smallskip}
\multicolumn{1}{l}{}
& \multicolumn{1}{|c}{DMS}
& \multicolumn{2}{c}{$z_{phot}$}
& \multicolumn{2}{c}{sim}
& \multicolumn{2}{|c}{HDF$_B$}
& \multicolumn{2}{c}{sim}
& \multicolumn{2}{c}{Parkes}
& \multicolumn{2}{c}{sim}\\
&&99\%&no $\chi^2$&99\%&no $\chi^2$&99\%&no $\chi^2$&99\%&no
$\chi^2$&99\%&no $\chi^2$&99\%&no $\chi^2$\\
\hline
\hline\noalign{\smallskip}
C&0.80&0.88&0.88&0.79&0.86&1.00&1.00&0.74&0.76&0.83&0.97&0.52&0.81\\
CR&0.23&0.27&0.24&0.20&0.21\\
E&0.18&0.24&0.21&0.16&0.18\\
\hline
\hline\noalign{\smallskip}
\end{tabular}
\end{table*}

In order to estimate the contamination due to compact galaxies,
we have added to the template library the spectra of blue starburst
galaxies, obtained from the public model Starburst99 code by Leitherer
et al. (1999), as well as a delta-burst model taken from the 
new Bruzual \& Charlot evolutionary code (GISSEL98, Bruzual \&
Charlot 1993), and a set of 4 empirical SEDs compiled by Coleman, Wu
and Weedman (1980) (hereafter CWW) to represent the local population
of galaxies (E, Sbc, Scd and Im). Spectra from the Starburst99 code
were selected with solar metallicity and Salpeter IMF, taking different
ages for an instantaneous burst ranging from 1 to 20 Myr. In the case
of the GISSEL98 delta-burst, the IMF is that of Miller \& Scalo (1979),
with solar metallicity and 51 different ages ranging from 0.001 to 20
Gyr.  CWW spectra were extended to wavelengths $\lambda \le
1400$\,\AA\ and $\lambda \ge 10000$\,\AA\ using the equivalent GISSEL98
spectra.

When applying the method to the DMS sample of 3523 stellar objects (we
remind here that we only use objects with photometry in all 6 filters), using 
the template library extended with galaxy spectra, we find that 
$\sim 80\%$ of them are better fitted by 
stellar or quasar templates rather than by galaxy spectra. This fraction is 
only 56\% for the subsample of 97 objects spectroscopically
identified as stars. Among the subsample of the 51 objects 
spectroscopically identified as quasars, there are 38\%
that show a better fit to the galaxy spectra, and 80\% among them
correspond in particular to genuine young starbursts according to the
best-fit spectra, with ages ranging between 1 and 20 Myrs. 
The result is different when the same extended template library is
applied to the 34 B $\le 22.3$ DMS CNELGs.
In this case, 85\% of the sample (29 objects) is well
fitted by galaxy spectra rather than quasar or stellar templates. The
fraction of best-fit galaxy templates corresponding to young starbursts
is $\sim 2/3$. Generally speaking, a typical fraction between $1/4$ and $1/3$ 
of the quasar candidates is
also compatible with young starbursts, and CNELGs are hardly selected as
quasars, when supplementary galaxy templates are used. 

To summarize, adding star-bursting and early-type galaxy
templates does not give a better solution to the problem of selecting
quasar candidates: if star-burst candidates are included in the list,
the number of candidates is multiplied by a factor of 4 to 5. On the
other side, objects like CNELGs will not be mixed with the quasar
candidates but we take the risk to miss $\sim 35\%$ of the quasars,
identified as star-bursting galaxies. However, for the purposes of a
survey, all objects will be tested through the same, general pipeline.
Some objects (like quasars and compact blue galaxies) 
will inevitably end up as candidates in more than one list, depending
on the different selection criteria.

\subsubsection{Quasar Photometric Redshifts}
\label{zphotreal}

\begin{figure*}[ht]
\centerline{
\psfig{figure=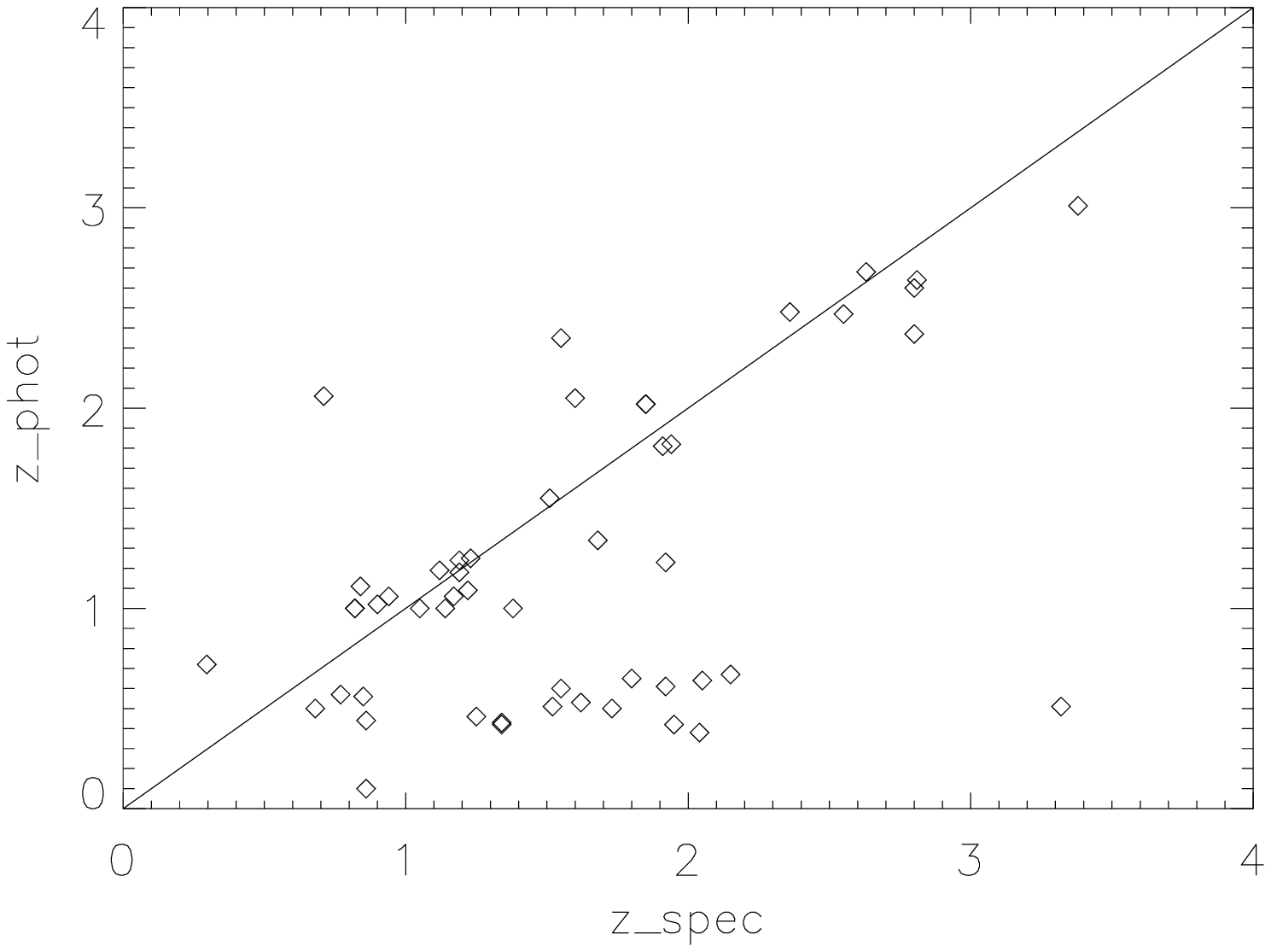,height=7cm,width=8cm}
\psfig{figure=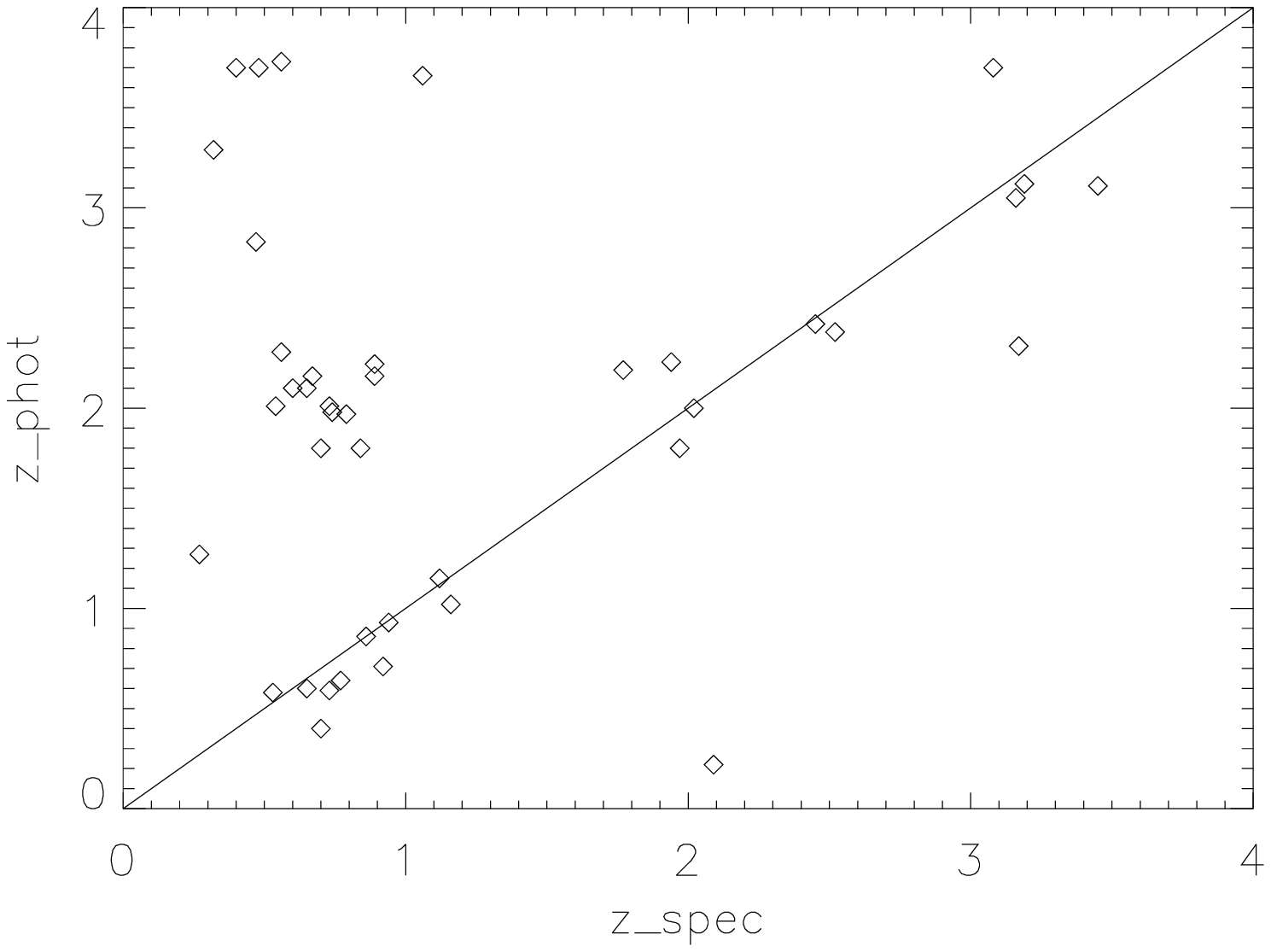,height=7cm,width=8cm}}
\caption{Photometric redshifts versus spectroscopic redshifts for the 51
quasars of the DMS sample (left) and for the new sample of 46 
spectroscopically confirmed quasars (Hall et al., 2000) (right).}
\label{zmzpps}
\end{figure*}

Figure \ref{zmzpps} left presents the $z_{phot}$ versus $z_{spec}$ for the
51 DMS quasars. For 23 of them, $|z_{phot}-z_{spec}| \le 0.1$. We note
an important dispersion for redshifts lower than typically 2.5: quasars
with a $z_{spec} \in [1,2]$ are better fitted by templates of quasars
with $z \in [0,1]$. This is also the tendency arising from simulations
and shown on figure 
\ref{zphotsimps} (upper right), for the filter combination UBVRI.

Figure \ref{zmzpps} right presents the $z_{phot}$ versus $z_{spec}$
for a new sample of 46 spectroscopically identified quasars (Hall et al.,
2000). 17 of them have $|z_{phot}-z_{spec}| \le 0.1$.
This time we note that quasars with a $z_{spec} \in [0,1]$ are better
fitted by quasars with $z \in [1,2]$, and this trend is also seen on the
simulations (figure \ref{zphotsimps} upper right) for the same set of
filters.

\subsection{HDF$_{Bright}$ and Parkes}

The {\it completeness} has also been tested 
on two recently published quasar
catalogues. Liu et al. (1999) published a list of 30 spectroscopically
confirmed quasars with their photometry in B, V and R. These objects
have $17.6 \le B \le 21.0$ and $0.44 \le z \le 2.98$, and were selected
on a one square degree field centered on the HDF North. They were
firstly identified as quasar candidates by their (B-V) (V-R) colors, along
with 31 other candidates that, after the spectroscopy, turned out to
be stars.

With our method we identified all 30 quasars as quasar candidates. However,
the limited number of filters did not allow us to reduce the number of
stars selected as quasar candidates, 30 out of 31 were in our own list.

We also applied our method to the Parkes subsample published by Francis
et al. (1999). Only point-like objects, with the photometry 
in all 7 filters (B, V, R, I, J, H, K$_n$) were kept, thus 129 out of the 157
quasars of the initial sample remain in our catalogue. At a confidence
level of 99\%, 107 among these objects were classified as quasar candidates
({\it completeness}=0.83). If no $\chi ^2$ selection is applied, the
{\it completeness} is equal to 0.97.

Columns 7 to 14 in Table \ref{efficiency} summarize these results.
Simulation results are also given for comparison. Note that in both
cases, the {\it completeness} for the real samples is found to be even 
higher than expected according to the simulations.

{\it Confirmation rate} and {\it efficiency} cannot be computed for
these two samples, since the entire point-like object catalogues 
(including stars) are not available.

\subsection{EIS-wide (patch B)}

An attempt to apply our method to the EIS--wide (patch B) data was made.
The results are approximatively the same as those
obtained in multicolor procedures.
This was somewhat expected: the multicolor approach
proposed here, when applied to a 3 filter data set, is equivalent to a
color--selection technique. Note that 3 filters have
been used to carry out this survey: B, V and I. On a (B-V) (V-I)
diagram, quasar evolutionary tracks up to a redshift of $\sim 3.3$
and the Main Sequence stars mix together. 

\section{Discussion}
\label{discuss}

The quasar multicolor selection technique presented in this paper has
some advantages as compared to the ``traditional'' color-color methods.
First, it has the advantage of selecting quasar candidates even at
redshifts where quasar and stellar colors are very much alike (2.5 -- 3.0).
However, the {\it efficiency} varies with the redshift. Independently of
the filter set, some quasars with redshifts lying in the interval [2.5,3]
will be skipped. Furthermore, the number of quasar candidates with
estimated $z_{phot}$ within this range will be far higher in comparison
to other redshift bins.  The set of filters used plays an important role 
in these results. Adding a filter
is not necessarily equivalent to adding information and does not always
improve the results, but imposes more stringent constraints by
increasing the number of degrees of freedom. For this reason, a $\chi^2$
selection does not improve the results in all cases.

This technique gives also an estimate of the (photometric) redshift 
of the candidates, and it is thus able to improve the spectroscopic
follow up of a given sample. In some cases, which depend (again) on the
filter combination and the redshift of the objects, this determination
can be accurate ($|z_{phot} - z_{spec}| \le 0.1$). In most cases, quasars 
at redshift $z \le 3$ will not be misidentified at $z \ge 4$ and vice
versa, therefore the spectroscopy can be reliably targeted towards
the optical or the infrared bands.

For the purposes of a survey aiming in assembling a quasar catalogue
with a minimum cost, adding star-bursting and early-type galaxy
templates does not appear as a priority. On the contrary, this addition
will cause either the increase of the candidates by a factor of almost
5 (if star-burst candidates are considered as possible quasars)
or the decrease of the {\it completeness} $\sim 40\%$ (because an
important fraction of real quasars will be better fitted by
star-burst galaxies templates). For practical purposes (e.g the VIRMOS
survey), when all the
sources examined through the same pipeline, a fraction of objects
identified as quasars will be also found within the ``starburst''
sample, but this does not affect the present selection.

The same method can be extended towards 
different directions. First, it can be adapted to particular objects 
like very red quasars, by adding the suited
templates. However, this has to be done thoroughly, to prevent from a
possible and undesirable increase of the quasar candidates. It can also
be used on spatially resolved objects, like Seyfert galaxies, because by
construction, the software is independent of the morphology of the
objects. This can be done by adding a stellar component to the AGN
simulated spectra already used. Last but not least, it can be extended to very
high redshifts (higher than 6), after an appropriate modeling of the 
intergalactic absorption.

As already mentioned, this software has been developed for the needs
of preparing the quasars/AGN sample assembled in the future
VIRMOS survey. However, the results and conclusions of this paper 
apply to all multicolor surveys in general.
A cross-identification between optical and X-ray sources can be
extremely useful, when X-ray data are available,
and can increase the {\it efficiency}, as quasar candidates with a 
compact X-ray counterpart will have an even higher chance of
being real quasars.


\begin{acknowledgements}
We would like to thank M. Bolzonella, J. P. Picat and M. Elvis 
for useful discussions. We would also like to thank the referee, Dr. P.
Hall for his very useful comments that led to the improvement of our
work.
\end{acknowledgements}


\begin{thebibliography}{}

\bibitem[Bolzonella et al., \- 2000]{BMP}
Bolzonella, M., Miralles, J. M. \& Pell\'o, R., 2000, submitted to A\&A,
pre-print astro-ph/0003380
\bibitem[Boyle et al., \- 1991]{boyle91} 
Boyle  B.J.; Jones, L.R., Shanks, T., 1991, MNRAS, 251, 482 
\bibitem[Bruzual \& Charlot, \- 1993]{BC}
Bruzual, G., Charlot, S. 1993, ApJ 405, 538
\bibitem[CUQS]{cuqs}
CUQS: http://sa1.star.uclan.ac.uk/~prn/cuqs.html
\bibitem[Cohen et al., \- 1999]{cohen}
Cohen, J., Hogg, D., Blandford, R., Cowie, L., Hu, E., Songaila, A.,
Shopbell, P. \& Richberg, K., 1999, accepted for publication
in ApJ, astro-ph/9912048
\bibitem[Coleman et al., 1980]{Cea}
Coleman, D.G., Wu, C.C., Weedman, D.W. 1980, ApJS 43, 393
\bibitem[Fan, X. 1999]{fan}
Fan, X., 1999, AJ, 117, 2528
\bibitem[Glazebrook et al., \- 1995]{Glaz}
Glazebrook, K., Ellis, R., Colless, M., Broadhurst, T., Allington-Smith,
J. \& Tanvir, N., 1995, MNRAS, 273, 157
\bibitem[Hall et al., \- 1996a]{hall} 
Hall, P. B., Osmer, P. S., Green, R. F., Porter, A. C., Warren, S. J. 
1996, ApJSS, 104, 185
\bibitem[Hall et al., \- 1996b]{hall2}
Hall, P. B., Osmer, P. S., Green, R. F., Porter, A. C., Warren, S. J. 
1996, ApJ, 462, 614  
\bibitem[Hartwick \& Schade, \- 1990]{HS}
Hartwick F. \& Schade D., 1990, ARA\&A, 28, 437 
\bibitem[Jarvis\- \& \- Mac\- Alpine,\- 1998]{jarvis} 
Jarvis, R.M. \&  MacAlpine, G.M., 1998, AJ, 116, 2624 
\bibitem[Kennefick\- et\- al., \- 1997]{kenn}
Kennefick J.D., Osmer, P.S., Hall P.B., Green, R.F., 1997, AJ, 114, 2269
\bibitem[Krisciunas et al.,\- 1998]{krisciunas} 
Krisciunas K., Margon, B., Szkody, P., 1998, PASP, 110, 1342
\bibitem[Leitherer et al., 1999]{Lea}
Leitherer, C., Schaerer, D., Goldader, J. D., Delgado, R. M., Robert, 
C., Kune, D. F., de Mello, D.F., Devost, D., Heckman, T. M., 1999,
ApJS 123, 3
\bibitem[Madau,\- 1995]{Mad}
Madau P., 1995, ApJ, 441, 18
\bibitem[Miller \& Scalo, \- 1979]{MS}
Miller, G.E., Scalo, J.M. 1979, ApJS 41, 513
\bibitem[Miller \& Mitchell,\- 1988]{mimi}
Miller, L. \& Mitchell, P. S., 1988, 
ASP Conference Series, Volume 2, Proceedings
of a Workshop on Optical Surveys for Quasars, (San Francisco: ASP),
ed. by Patrick Osmer and M.M. Phillips, p. 114
\bibitem[Miralles \& Pell\'o,\- 1998]{JMM-R}
Miralles J.M. \& Pell\'o, R., 1998, astro-ph/9801062  
\bibitem[Newberg \& \- Yanny,\- 1997]{newberg}
Newberg, H.J. \&  Yanny, B., 1997, ApJSS, 113, 89
\bibitem[Osmer et al., \- 1998]{osmer}
Osmer, P., Kennefick J.D., Hall, P.B., Green, R.F., 1998, ApJS, 119, 189, and
http://www.astronomy.ohio-state.edu/~posmer/DMS/
\bibitem[Pell\'o et al., \- 1999]{pello2}
Pell\'o, R., Kneib, J.-P., Bolzonella, M., Miralles, J.-M., 1999, 
Proceedings of the ``Photometric Redshifts and High Redshift Galaxies'', 
Pasadena, PASP Conf. Ser. 191, p 247.
[ astro-ph: 9907054]
\bibitem[Peri, Iovino \& Hickson, 1997]{PIH}
Peri, F., Iovino, A. \& Hickson, P., 1997, ``Quasar Detection using 
Multi-Narrow-Band Photometry'', in {\it Science with Liquid 
Mirrors Telescopes}, Proceedings of the Marseille Meeting on LMTs, April
1997, in press.
\bibitem[Peterson, 1997]{Peterson}
Peterson, B. M., 1997, in {\it An introduction to active galactic
nuclei}, Cambridge University Press 
\bibitem[Pickles, 1998]{Pickles}
Pickles A. J., 1998, PASP, 110, 863
\bibitem[Prandoni et al.,\- 1998]{EIS III} 
Prandoni I., Wichmann R., da Costa L., Benoist, C., Mendez, R., Nonino, M.,
Olsen, L. F., Wicenec, A., Zaggia, S.,
Bertin, E., Deul, E., Erben, T., Guarnieri, M. D., Hook, I., Hook, R.,
Scodeggio, M., Slijkhuis, R., 1999, A\&A, 345, 448
\bibitem[Robin et al., 1995]{robin}
Robin, A., Haywood, M., Gazelle, F., Bienaym\'e, O., Cr\'ez\'e, M.,
Oblak, E., Guglielmo, F. 1995, ``Model of Stellar Population Synthesis of
the Galaxy'', http://WWW.obs-besancon.fr/www/modele/modele\_ang.html
\bibitem[Sloan Digital Sky Survey]{sdss} 
http://www.sdss.org/news/releases/19981208.qso.html
\bibitem[Wang et al., 1998]{wang}
 Wang, T. G., Lu, Y. J. \& Zhou, Y. Y., 1998, ApJ, 493, 1
\bibitem[Warren et al.,\- 1994]{warren} 
Warren S.J., Hewett, P.C., Osmer, P.S. 1994, ApJ, 421, 412 
\bibitem[Wolf et al.,\- 1998]{wolf}
Wolf, C., Meisenheimer, K., Roeser, H.-J., 1999, A\&A, 343, 399

\end{thebibliography}
\end{document}